\documentclass[aps,pre,twocolumn,groupedaddress]{revtex4-1}

\usepackage{graphicx,url,hyperref}
\usepackage{amsmath,amssymb}

\def\sign{\mathop{\rm sign}\nolimits}

\newcommand{\eref}[1]{(\ref{e:#1})}
\newcommand{\fref}[1]{Fig.~\ref{f:#1}}

\newcommand{\ct}[1]{\cite{#1}}

\begin{document}

\title{Effects of mixing in threshold models of social behavior}

\author{Andrei R. Akhmetzhanov}
\email{aakhmetz@um2.fr}
\altaffiliation[Currently at ]{Institute of Evolutionary Sciences, University of Montpellier II, Montpellier 34095, France}
\author{Lee Worden}
\email{worden.lee@gmail.com}
\altaffiliation[Also at ]{UC Berkeley and San Francisco Art Institute, San Francisco, California, USA}
\author{Jonathan Dushoff}
\email[Corresponding author]{dushoff@mcmaster.ca}
\affiliation{Theoretical Biology Laboratory, Department of Biology, McMaster University, Hamilton, Ontario L8S4K1, Canada}

\date{\today}

\begin{abstract}
We consider the dynamics of an extension of the influential Granovetter model of social behavior, where individuals are affected by their personal preferences and observation of the neighbors' behavior.  Individuals are arranged in a network (usually, the square lattice) and each has a state and a fixed threshold for behavior changes.  We simulate the system asynchronously either by picking a random individual and either update its state or exchange it with another randomly chosen individual (mixing). We describe the dynamics analytically in the fast-mixing limit by using the mean-field approximation and investigate it mainly numerically in case of a finite mixing. We show that the dynamics converge to a manifold in state space, which determines the possible equilibria, and show how to estimate the projection of manifold by using simulated trajectories, emitted from different initial points.

We show that the effects of considering the network can be decomposed into finite-neighborhood effects, and finite-mixing-rate effects, which have qualitatively similar effects. Both of these effects increase the tendency of the system to move from a less-desired equilibrium to the ``ground state". Our findings can be used to probe shifts in behavioral norms and have implications for the role of information flow in determining when social norms that have become unpopular (such as foot binding or female genital cutting) persist or vanish.
\end{abstract}


\maketitle

\section{Introduction}

In this paper, we investigate a simple model of behavior, the threshold model (TM) \ct{Schelling1971,Granovetter1978}. It consists of $N$ individuals arranged in a network. Each individual, described by a state variable $s_i$ ($i=1,\ldots,N$), has either adopted or rejected the behavior in question and has a tendency to switch to adopting (rejecting) if the proportion of individuals in its neighborhood adopting the behavior is greater (less) than its (constant) threshold $T_i$. Individuals are chosen at random to be ``updated" -- i.e., to consider, and possibly change (``flip") their state.  We make an analogy with physics by thinking of the individual's state as a `spin' with value $+1$ ($-1$) for those who adopt (reject) the behavior.

Threshold models are relevant to questions of how patterns of behavior persist, even when attitudes change, and how these patterns can sometimes change rapidly.  A currently relevant example is the practice of \emph{female genital cutting} (FGC), which goes back at least to ancient Egypt \ct{Kennedy2009}. Despite a public health consensus that the practice is harmful \ct{WHO2008}, traditional practice remains widespread in various societies \ct{TagEldin2008,UNICEF2010}.  A similar example is the Chinese practice of footbinding, which was widely practiced for hundreds of years, before disappearing rapidly \ct{Mackie1996}.  These practices can be considered in the context of the theory of ``social norms", behaviors which individuals prefer to follow, \emph{given that they think that others will conform, and that others expect them to conform} \ct{Bicchieri2006}. 

Many similar individual-based models are also based on individuals making binary choices \ct{SanMiguel2005,Castellano2009}. Usually, the voter model \ct{Liggett1999} is associated with imitation process, since a randomly chosen individual adopts the behavior of one of its neighbors. In this sense, the TM puts the social pressure in the framework: the adoption or rejection of the behavior by an individual depends on the current level of adoption in its neighborhood \ct{Vilone2012}. The majority rule model (MR, see~\ct{Krapivsky2003}) is a special case of the TM, since all thresholds are one half (the randomly chosen individual tends to flip if the \emph{majority} of its neighbors have opposite spin). 

It has been shown that the majority rule model can be described by the classical Ising model with zero external magnetic field~\ct{Krapivsky2003} and that the general TM can be described as a random-field Ising model (RFIM)~\ct{Barra2012}. The study of RFIMs in physics often focuses on critical temperature phenomena \ct{Dorogovtsev2008} or metastable states and hysteresis loop phenomena at zero temperature \ct{Sethna1993,Rosinberg2009b}. Instead of using the notion of the thermodynamic temperature, where individuals probabilistically flip in a non-preferred direction, see, for example, \ct{Brock2001,Malarz2011}, we chose to set the thermodynamic temperature to zero and study the effects of mixing on the dynamics.

We simulate our model on a two-dimensional lattice, with global mixing.  We implement mixing by allowing individuals to exchange places within the network at rate $\mu$ (relative to the update rate). The importance of mixing in sociological and ecological studies has been  demonstrated in other contexts \ct{Levin1974,Blasius1999,Agliari2006,Reichenbach2007}.  Introducing global mixing on a two-dimensional lattice is similar in concept to using a ``small-world" network \ct{Watts1998}.  Both cases have regular connections, and random global connections -- the difference is that we implement random global connections by switching individuals.



We simulate behavior change by either choosing an individual at random to update or mixing two individuals in each step of the simulation. Mixing consists of exchanging two randomly chosen individuals rather than updating in a given simulation step, with probability $\mu/(1+\mu)$, so that we have an average of $\mu$ switches per update.  We increment the clock by $1/N$ per update event.  This gives us update events at rate 1 per individual and mixing events at rate $\mu$ per individual.  When we mix individuals, we exchange their states and thresholds, leaving the network otherwise unchanged.  A synchronous process or a pure Poisson process would be expected to give qualitatively similar results, but this asynchronous process is simpler to simulate, and can be analyzed using an ordinary differential equation (ODE) framework derived from master equations. 

Most analytical results in the field of TMs/RFIMs are obtained using mean-field approximations \ct{Dominicis2006,Krapivsky2010}. This can be achieved either by considering a complete network (where every node is a neighbor of every other node), or by setting the mixing rate $\mu\gg1$.

The intermediate mixing case $\mu\sim1$ is not so easily treated. If we write equations for the moments of different order for the distribution of states among individuals, we get a hierarchical system of coupled equations~\ct{Krapivsky2010}. There are then various methods to ``close" the system by approximating higher moments in terms of lower moments~\ct{Bolker1999,Murrell2004,Murrell2009}.

In \ct{Toral2007}, the authors considered a MR model, and concluded that the behavior of their system resembles the movement of a Brownian particle in a potential field that is unknown \emph{a priori}. We describe such an ``effective potential'' function for our threshold model and calculate the analytical potential form for the mean-field version of the model.

We can use the effective potential to provide an additional perspective on the dynamical properties of the system.  The bifurcation where the system changes from having one stable equilibrium to two, for example, corresponds to a change from a single-welled to a double-welled effective potential function. In terms of an Ising-like model, this would correspond to a phase transition of the first order. This bifurcation is relevant from a sociological point of view, since a transformation from a potential consisting of two wells to a potential consisting of one well, due to a change of mixing rate, could give rise to sudden abandonment or adoption of a social norm. In contrast, if such transformation does not occur, even when one well is much deeper than the other, this might help to explain why a human society sometimes continues to support a fairly unpopular social norm for many years~\ct{Bicchieri2006}.

In this paper we explore the dynamics of this system using a Gaussian threshold distribution. It is not necessary to truncate the distribution, since we can simply assume that if the threshold is ${<}0$ (or ${>}1$), an individual will always be updated to its preferred state independent of its neighborhood configuration. We expect other flexible distributions to give similar qualitative results. For example, even uniform distributions show the same basic bifurcations that we explore here \ct{Granovetter1978}.  We have also tried simulations with bimodal superpositions of Gaussians, and again seen qualitatively similar results. 

Here is how we organize the rest of the paper. First we consider the mean-field dynamics which are given by the fast-mixing and large-scale ($N\gg1$) limits. Then we consider the intermediate mixing rates and state the main results of our paper: we discuss the bifurcation phenomena found in the TM and we demonstrate the appearance of a  manifold in the dynamics that is approached by any trajectory of the TM. Finally, we interpret our main results from a sociological point of view, and draw conclusions.

\section{Mean-field approximation}

First, consider the case in which the neighborhood size is so large that each spin is connected with all other spins in the network. In this case, we recover Granovetter's threshold model for collective behavior \ct{Granovetter1978}, with dynamics in which the probability of a spin being updated from minus to plus is given by $\mathbb P(\uparrow|y)=(1-y)F(y)$, where $y$ denotes the proportion of plus spins and $F(\cdot)$ is the cumulative distribution function of the thresholds' PDF $f(x)$:
\begin{equation}\label{e:F}
F(x)=\int_{-\infty}^x f(\xi)\mathrm d\xi\;.
\end{equation}
The probability of a spin being flipped in opposite direction from plus to minus is $\mathbb P(\downarrow|y)=y(1-F(y))$. In this case, master equations can be written in terms of the probability function $p(y_k,t)$ ($y_k=k/N$), which provides the probability to find the TM in a state with $k$ spins in a plus state and $N-k$ spins in a minus state at a given moment of time $t$:
\begin{multline}\label{e:MEq1}
\frac{\mathrm dp(y_k,t)}{\mathrm dt} = \mathbb P(\uparrow|y_{k-1})p(y_{k-1},t)\\{}+\mathbb P(\downarrow|y_{k+1})p(y_{k+1},t)-\mathbb P(\updownarrow|y_k)p(y_k,t)\,, \\ k=0,\ldots,N\,,
\end{multline}
where $\mathbb P(\updownarrow|y_k)=\mathbb P(\uparrow|y_k)+\mathbb P(\downarrow|y_k)$.

Letting $N\rightarrow \infty$ and scaling the time as $t\rightarrow Nt$, we can treat the discrete variable $y_k$ as continuously changing $y\in[0,1]$ and transform \eref{MEq1} to the Hamilton-Jacobi equation, which is the first order partial differential equation:
\begin{equation}
\label{e:HJEq1}
\frac{\partial p(y,t)}{\partial t} = -\frac{\partial}{\partial y}[(F(y)-y)p(y,t)]\;.
\end{equation}
During such transformation, the diffusive terms, consisting of second order partial derivatives, vanish due to the large-scale limit $N\gg1$.

If the initial configuration is strictly defined such that $p(y,0) = \delta(y-y_0)$, where $\delta(\cdot)$ is the Dirac delta, the solution of \eref{HJEq1} is given by the following ODE, see p.~53--54 \ct{Gardiner2004}:
\begin{equation}\label{e:ODE1}
\dot y \equiv \frac{\mathrm dy}{\mathrm dt}=F(y)-y,\quad y(0)=y_0\;.
\end{equation}
The equilibria of this system are all values $y_*$ for which $F(y_*)=y_*$.

Notice that \eref{ODE1} can also be written in terms of the potential function: $V(x)=-\int^x(F(\xi)-\xi)\mathrm d\xi$, such that $\frac{\mathrm dy}{\mathrm dt}=-V'(y)$. Thus, the equilibrium points can be also defined by the extrema of $V(y)$. \medskip

\begin{figure}[!tb]
\hspace*{-.05\columnwidth}\centerline{\includegraphics[width=.75\columnwidth]{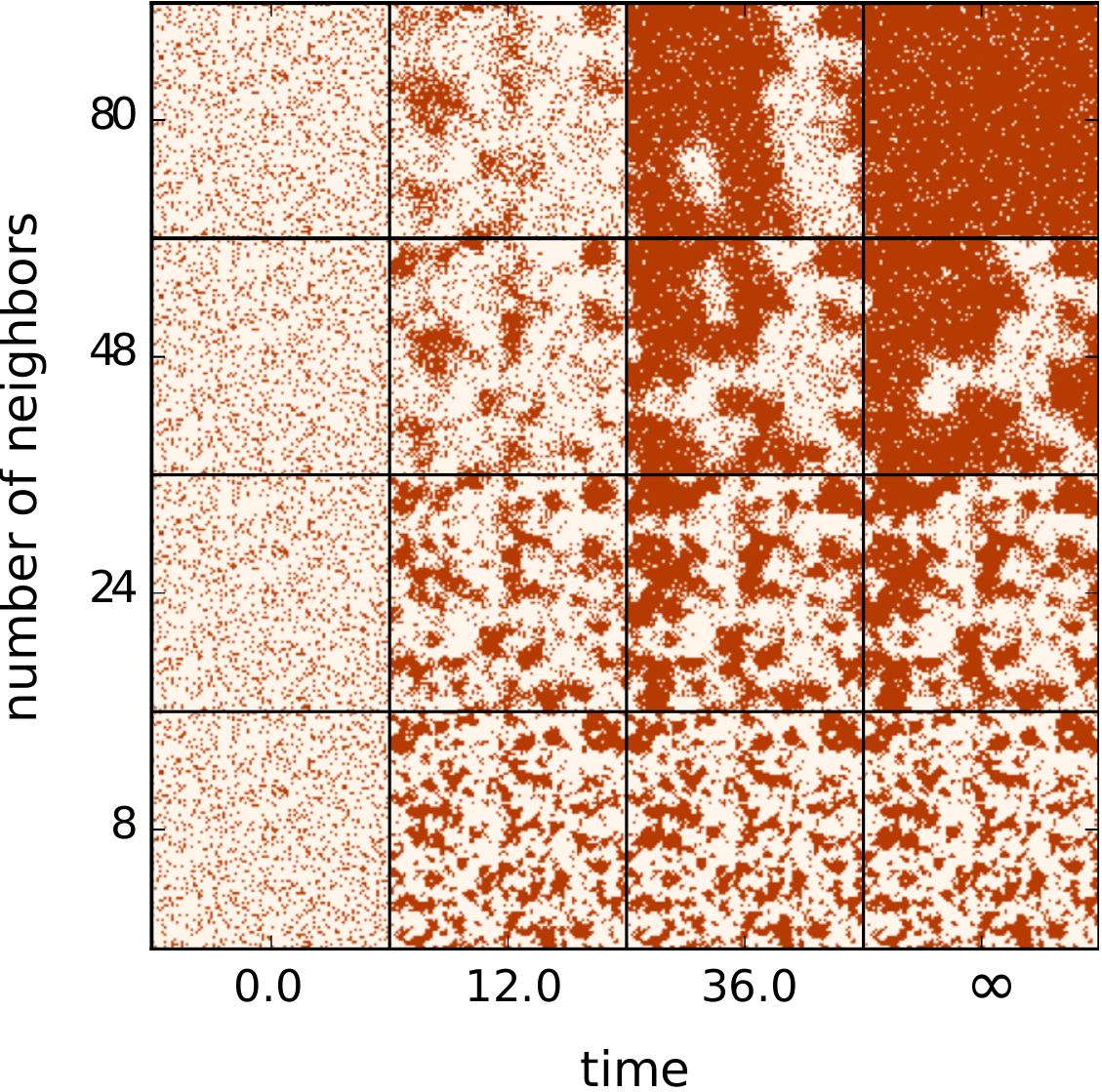}}
\caption{Simulations of the threshold model (TM) on a two-dimensional lattice of size $100^2$ with no mixing among individuals, and different neighbourhood sizes. The thresholds are normally distributed with the mean $0.45$ and standard deviation $0.3$. The initial pattern of thresholds and initial states is the same for all simulations shown. Activated individuals are shown in orange (black). The $\infty$-symbol denotes an equilibrium, which is reached in the TM.}
\label{f:tsp0}
\end{figure}

We now consider a case where each individual's updates depend on the states in a \emph{finite} neighborhood. First simulations of the TM on a two-dimen\-sio\-nal lattice with 8 nearest neighbors for each individual and with no mixing among them reveal complex patterns, see~\fref{tsp0}. This figure presents initial, intermediate and final states of the lattice for four different neighborhood sizes, but with identical initial distributions of states and thresholds. We see that increasing neighborhood size can shift the outcome of the system's dynamics from a low level of conformity to a very high level. Moreover, the equilibrium distribution preserves some noticeable clustering for small neighborhood sizes.

However, in the large-scale ($N\rightarrow\infty$), fast-mixing ($\mu\rightarrow\infty$) limit, the behavior of the TM can still be described analytically by a mean-field model, in which a spin and all its neighbors are chosen \emph{de novo} at each update event.

The probability that a randomly selected individual will choose to adopt is equal to the probability that the activation level of its randomly selected neighborhood exceeds its threshold.  In a regular network where each individual has $n$ neighbors, this is given by:
\begin{equation}\label{e:Fn}
F_n(y) = \sum_{k=0}^nF\! \left(\frac kn \right)C_n^k y^k(1-y)^{n-k}\,,
\end{equation}
where $C_n^k=\frac{n!}{k!(n-k)!}$ is a binomial coefficient (see~\ct{Barra2012}). Then \eref{MEq1}-\eref{ODE1} remain valid for this system once we substitute $F_n(y)$ for $F(y)$ in \eref{Fn}, and they give us the dynamics of the TM with finite neighborhood size in the mean-field approximation.  (This approach also works for networks of variable degree; if neighborhood sizes are distributed with probability density $\mathcal P(n)$, we average over the distribution to get $F_\mathcal P = \sum_{n=0}^\infty\mathcal P(n)F_n(y)$.)

\begin{figure}[!tb]
\hspace*{-.06\columnwidth}\includegraphics[width=0.75\columnwidth]{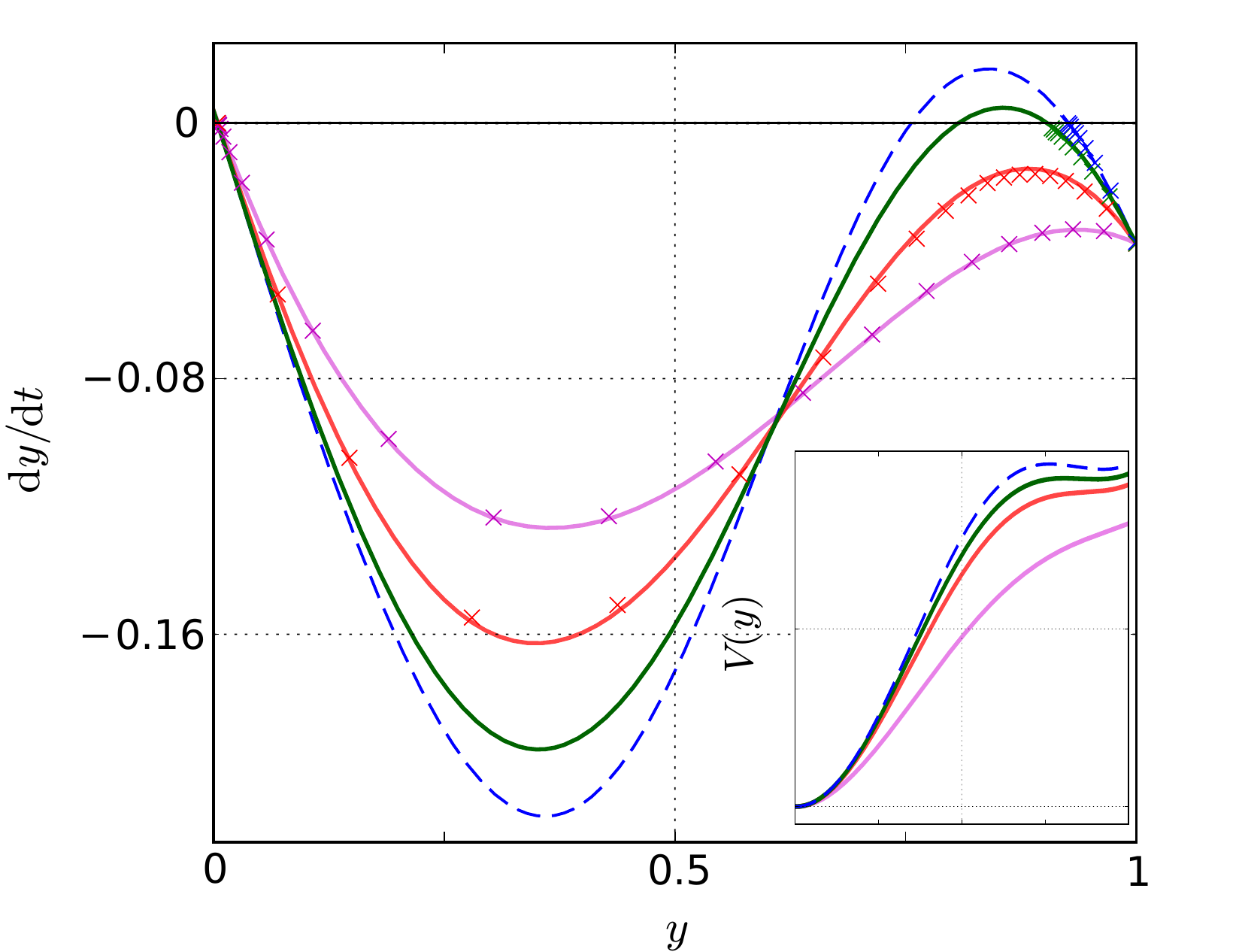}
\caption{(Color online) The mean-field dynamics of the TM in the phase plane $(y,\mathrm dy/\mathrm dt)$ for different neighborhood sizes: $n=4$ (magenta (light)), 12 (red (medium)) and 24 (green (dark)). The dashed curve corresponds to the case of infinitely large neighborhood size. The corresponding potential functions $V_n(y)$ are shown in the inset. The thresholds' PDF is Gaussian with the mean $0.6$ and standard deviation $0.225$. The crosses show the results of numerical simulations of the TM on a two-dimensional lattice of the size $100^2$ at mixing rate $\mu=4$ and $y(0)=1$}
\label{f:tsp1}
\end{figure}

\fref{tsp1} illustrates the difference between the functions \eref{Fn} and \eref{F} as well as the difference in the mean-field dynamics of the TM for different neighborhoods.  Parameters of the thresholds' PDF $f$ for \fref{tsp1} were chosen in such a way that there are two stable equilibria for large neighborhoods, and only one stable equilibrium for small neighborhoods. Different neighborhood sizes can lead to very different outcomes, even when distributions and initial conditions are the same. For example, in simulations starting with everybody adopting the behavior ($y(0)=1$), the TM reaches a high equilibrium (few individuals change), when neighborhood size is large, but a low equilibrium (almost everybody rejects the behavior) when neighborhood size is small. Note that simulations done on a two-dimensional lattice of size $100^2$ at mixing rate $\mu=4$  give a good approximation to the large-scale, fast-mixing limit in this case; later we will show that this is not true for smaller mixing rates, though.

In case of a Gaussian distribution for the thresholds, the curve $y'=F_n(y)$ has up to three crossings with the diagonal $y'=y$. If there is only one crossing with the diagonal, there exists a globally stable equilibrium $y_-\in[0,1]$. If there are three crossings, we have three equilibrium points, which we denote as $y_-$, $y_*$ and $y_+$, such that $y_-<y_*<y_+$. Two of them, $y_\pm$, are stable equilibria and one of them, $y_*$, is an unstable equilibrium.

We can define a potential, analogous to the mean-field case: $V_n(y)=\int^y(F_n(\xi)-\xi)\>\mathrm d\xi$.  Then $y_\pm$ are the minima and $y_*$ is the maximum of $V_n(y)$, see~\fref{tsp1} (inset).  The case of two crossings represents the bifurcation point between these two generic cases.

If the mean of the Gaussian distribution is not exactly $0.5$, the potential function $V_n(y)$ is asymmetric, with one well deeper than the other. Without loss of generality, we assume that the norm is intrinsically unpopular (i.e., the mean of the threshold distribution $>\!0.5$), so that the ``lower'' equilibrium $y_-$ corresponds to the deeper well, and the ``upper'' equilibrium and $y_+$ to the shallower well, when it exists.  These values refer to the case where $\mu \to \infty$.  For clarity, we will sometimes add $\infty$ as a superscript.

\section{Intermediate mixing}

Simulations show that reducing the mixing rate away from the fast-mixing limit has a similar qualitative effect to reducing neighborhood size (as seen in~\fref{tsp1}).  In the case where the mean-field system has one stable equilibrium, reducing mixing rates does not lead to a qualitative change in the dynamics. In the case where the mean-field system has two stable equilibria, as the mixing rate gets smaller we often find a bifurcation to a single equilibrium; i.e., the equilibrium with the shallower potential well disappears.

We consider a two-dimensional lattice, with initial activation level $y(0)$.  If we simulate, starting from a value between the two stable equilibria, the system will move to the upper equilibrium with probability $p_+$; otherwise it moves to the lower equilibrium.  The result depends on the random selection of thresholds, initial states and the order in which sites are updated.

\begin{figure}[!tb]
\hspace*{-.08\columnwidth}\includegraphics[width=.8\columnwidth]{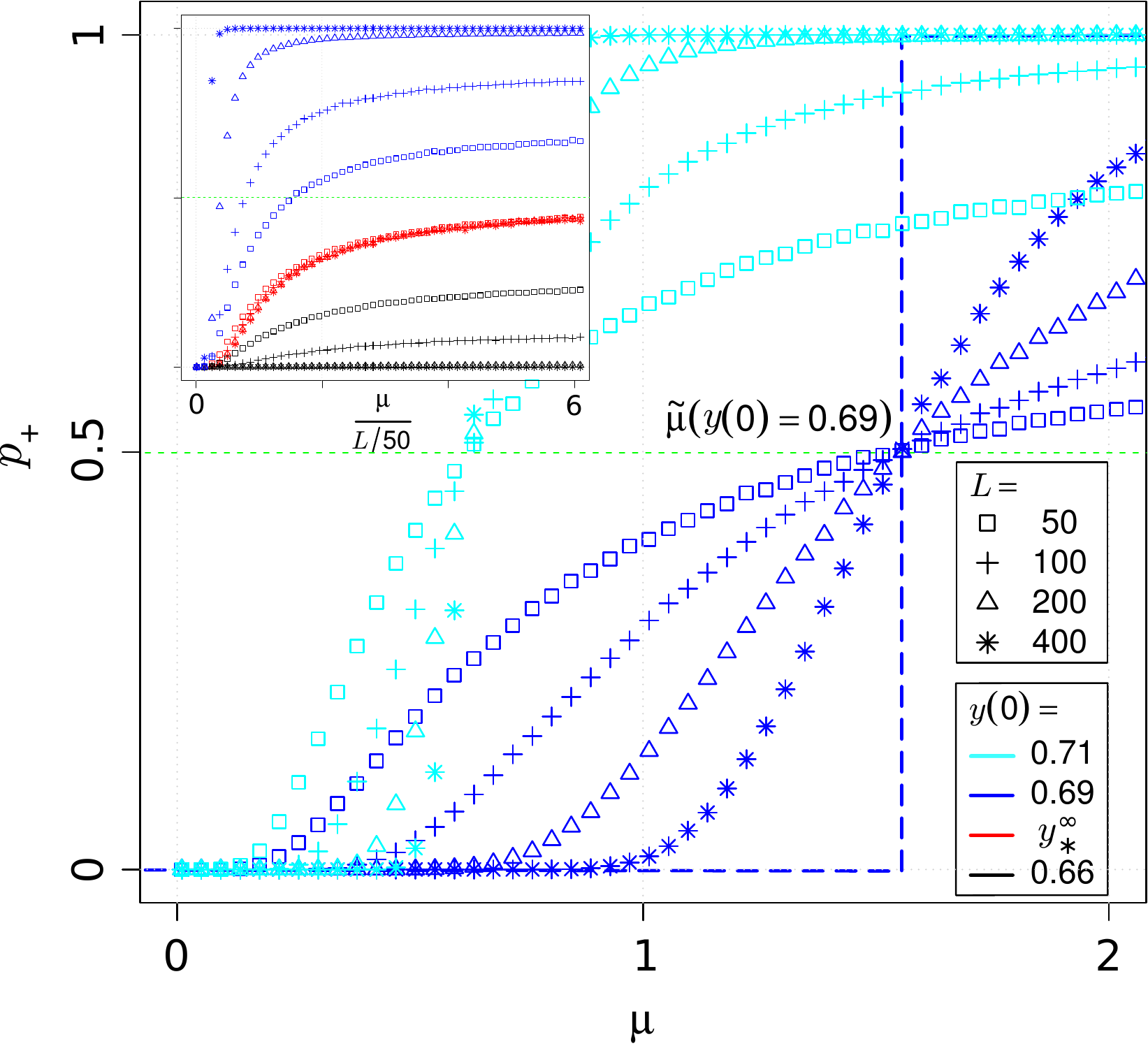}
\caption{(Color online) The probability $p_+$ that the TM will approach the upper equilibrium as a function of $\mu$ or scaled $\mu/L$~(the inset). The TM is on a 2d-lattice of size $N=L^2$ with the neighborhood $8$. The thresholds' PDF is Gaussian with the mean $0.55$ and standard deviation $0.225$. Different colors (shades) stand for different initial values $y(0)$, while the symbols stand for different values of $L$ (see the legend in the bottom right corner). In the main figure the points with $y(0)=0.71$ are shown in cyan (light), with $y(0)=0.69$ in blue (dark). In the inset overlapping points in the middle correspond to $y(0)=y_*^\infty$, the ones above them to $y(0)=0.69$, and the ones below them to $y(0)=0.66$. The bifurcation value $\tilde\mu(y(0)=0.69)\approx1.556$ is indicated. To estimate the probability, we performed $10^5$ simulations with different random initial individual states and thresholds, and update order}
\label{f:tsp4}
\end{figure}

\fref{tsp4} shows how the probability $p_+$ depends on the mixing rate $\mu$, using two different scaling approaches. In either case, as we move away from the fast-mixing case (from right to left on the figure), the system becomes increasingly certain to end in the deeper well, and eventually the shallower well disappears altogether.

The main picture of \fref{tsp4} shows the probability $p_+$ vs.\ $\mu$ for values of $y(0) > y_*^\infty$.  In this case the system stops in the shallow well for large values of $\mu$, and moves to the deeper well for smaller values. This transition becomes steeper as the size of the network increases. The curves for a given starting point intersect where $p_+=1/2$.  That is to say, for a given value of $\mu$, the value of $y(0)$ that falls ``in the middle" of the two wells -- so that the system is equally likely to go to either one -- does not change with lattice size.  We call this value  $y_{\times}^\mu$, because it is related to the unstable equilibrium $y_*^\mu$, but not equivalent (as we will see below). The dependence of $y_\times^\mu$ on $\mu$ has a hyperbolic shape and its minimal value $\bar\mu$ is reached at $y_\times^\mu=1$, which is shown in~\fref{y0_vs_muC}.

\begin{figure}[!tb]
\includegraphics[width=0.72\columnwidth]{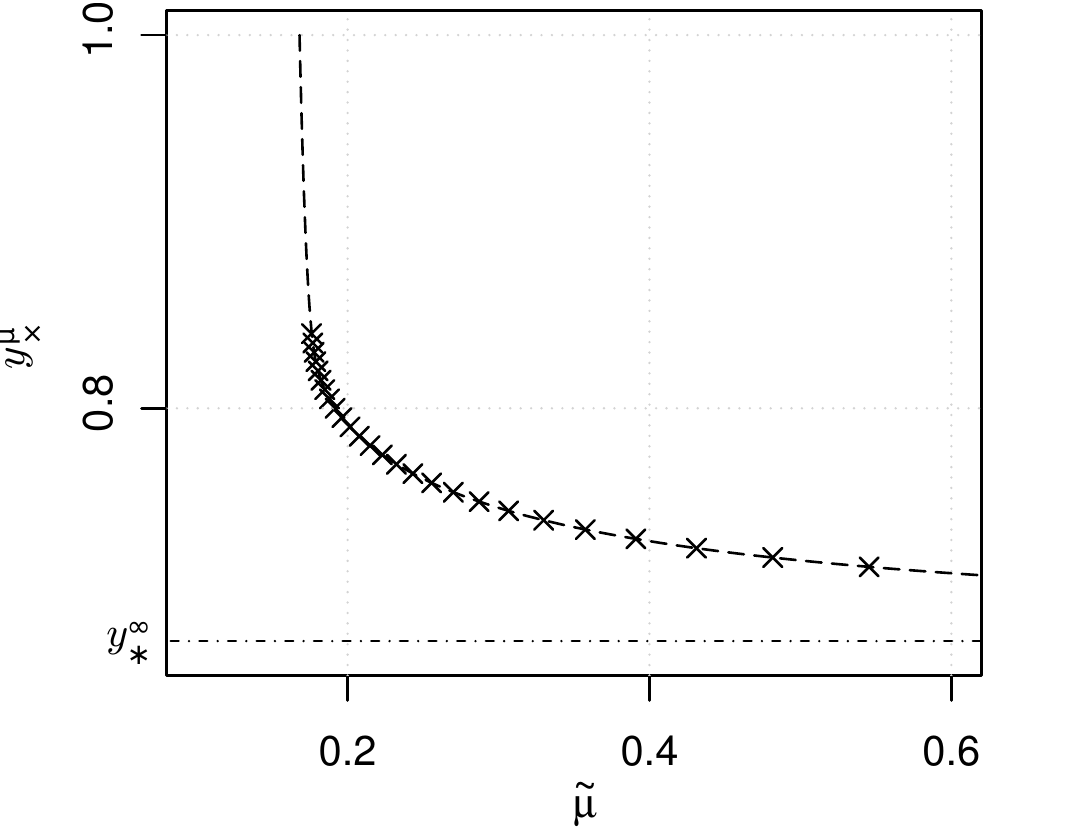}
\caption{The dependence of $y_\times^\mu$ on $\mu$. The TM is posed on a 2-d lattice of size $200^2$ with neighborhood $8$ and Gaussian distribution of thresholds with the mean $0.55$ and standard deviation $0.225$. The extrapolation curve, shown by the dashed line, gives $\bar\mu = \mu(y_\times^\mu=1.0)\approx 0.168$.}
\label{f:y0_vs_muC}
\end{figure}

The inset in \fref{tsp4} shows the same data, with $p_+$ plotted against a \emph{scaled} version of the mixing rate $\mu/L$ (where $L=N^{1/2}$ is the length of the two-dimensional lattice). A surprising pattern emerges. For $y(0) = y_*^\infty$, all of the curves approximately align onto a single curve, with $p_+\rightarrow1/2$ for $\mu\rightarrow\infty$, as we expect, since we are approaching the well-mixed case, where $y(0)$ is the unstable equilibrium. For other values of $y(0)$, the curves do not intersect in this scaling: instead, as $N$ gets larger, the system becomes less likely to ``switch" to the equilibrium on the other side of $y_*^\infty$.

If we visualize the trajectories in the phase subspace $(y,\mathrm dy/\mathrm dt)$,  we find that all of them approach the same curve $F_n^\mu$, shown in~\fref{SM1}, presumably because they are collapsing onto a lower-dimensional slow manifold. Thus, the behavior of the TM can be well-approximated by the ODE: ${\mathrm dy}/{\mathrm dt} = F_n^\mu(y)-y$, on some time interval $t\in[t_1,\infty)$, which is similar to \eref{ODE1}, where $F_n^\infty\equiv F_n$. The equilibrium points $y_*^\mu$ can be determined as $F_n^\mu(y_*^\mu)=y_*^\mu$ and the effective potential can be introduced by $V_n^\mu(y)=-\int^y(F_n^\mu(\xi)-\xi)\mathrm d\xi$. Thus, we can describe the behavior of the TM qualitatively, by studying the properties of the manifold-projection curve $F_n^\mu$.

\begin{figure}[!tb]
\hspace*{-.12\columnwidth}\includegraphics[width=.85\columnwidth]{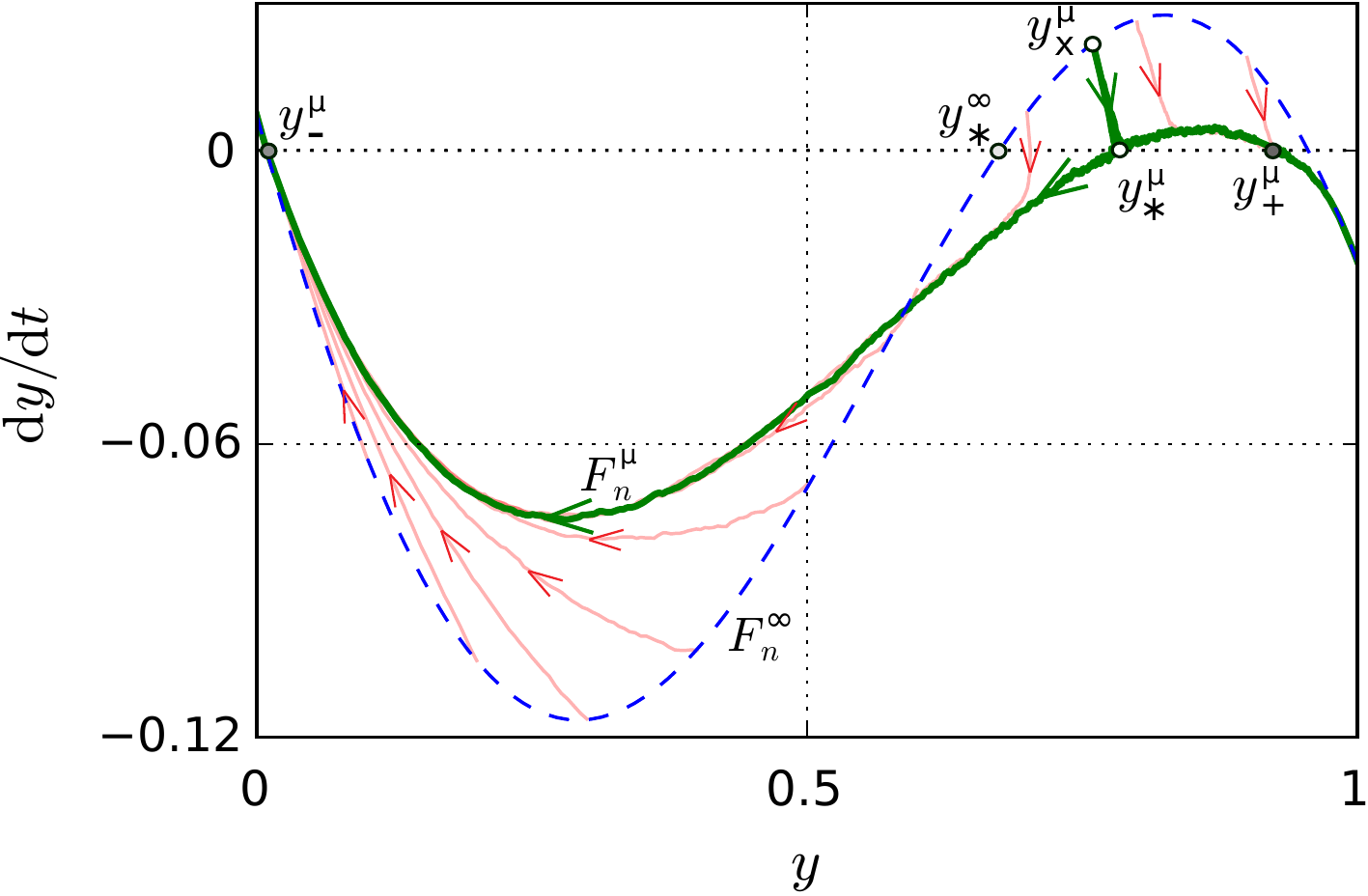}
\caption{(Color online) The projection curve $F_n^\mu$  on a two-dimensional lattice of size $800^2$ with neighborhood $8$ and the Gaussian distribution of thresholds (the mean 0.55 and standard deviation $0.225$ and mixing rate equals $0.278$. The trajectories, shown in red (thin) lines, start from $y(0)=i/10$ ($i=1,\ldots,9$), while the green (solid) curve consists of the trajectories initiated at $y(0)=0$, $y(0)=1$ and $y(0)=y_\times^\mu=0.76$. Initially, the states and thresholds are distributed randomly among individuals, hence all initial points fall along the curve $F_n(y)\equiv F_n^\infty(y)$ (dashed curve)}
\label{f:SM1}
\end{figure}

When $y$ approaches 0 or 1, mixing does not affect the dynamics.  Therefore, we can construct at least part of the curve $F_n^\mu$ (for any value of $\mu$) by simulating trajectories starting from $y(0)=1$ and $y(0)=0$. When there is only one equilibrium in $[0,1]$, this method generates the whole projection curve. When there are two stable equilibria, this method generates only the part ``outside" them.  Completing the curve requires that we start simulations from one or more intermediate initial points $y(0)\in(0,1)$. In fact, we need only one additional starting point, which is precisely $y_{\times}^\mu$, since any trajectory initiated at that point goes through the point $y_*^\mu$ on the curve $F_n^\mu$ to the upper or lower equilibrium with equal probability one half.

Note that if we take the minimal value $\bar\mu$ of the mixing rate, which can defined using~\fref{y0_vs_muC}, $F_n^\mu$ will have two fixed points, one of them will be a double root of $F_n^\mu(y_*^\mu)=y_*^\mu$, which corresponds to the bifurcation, described above.

\subsection*{Transition times}

\begin{figure}[!tb]
\hspace*{-.05\columnwidth}\includegraphics[width=0.85\columnwidth]{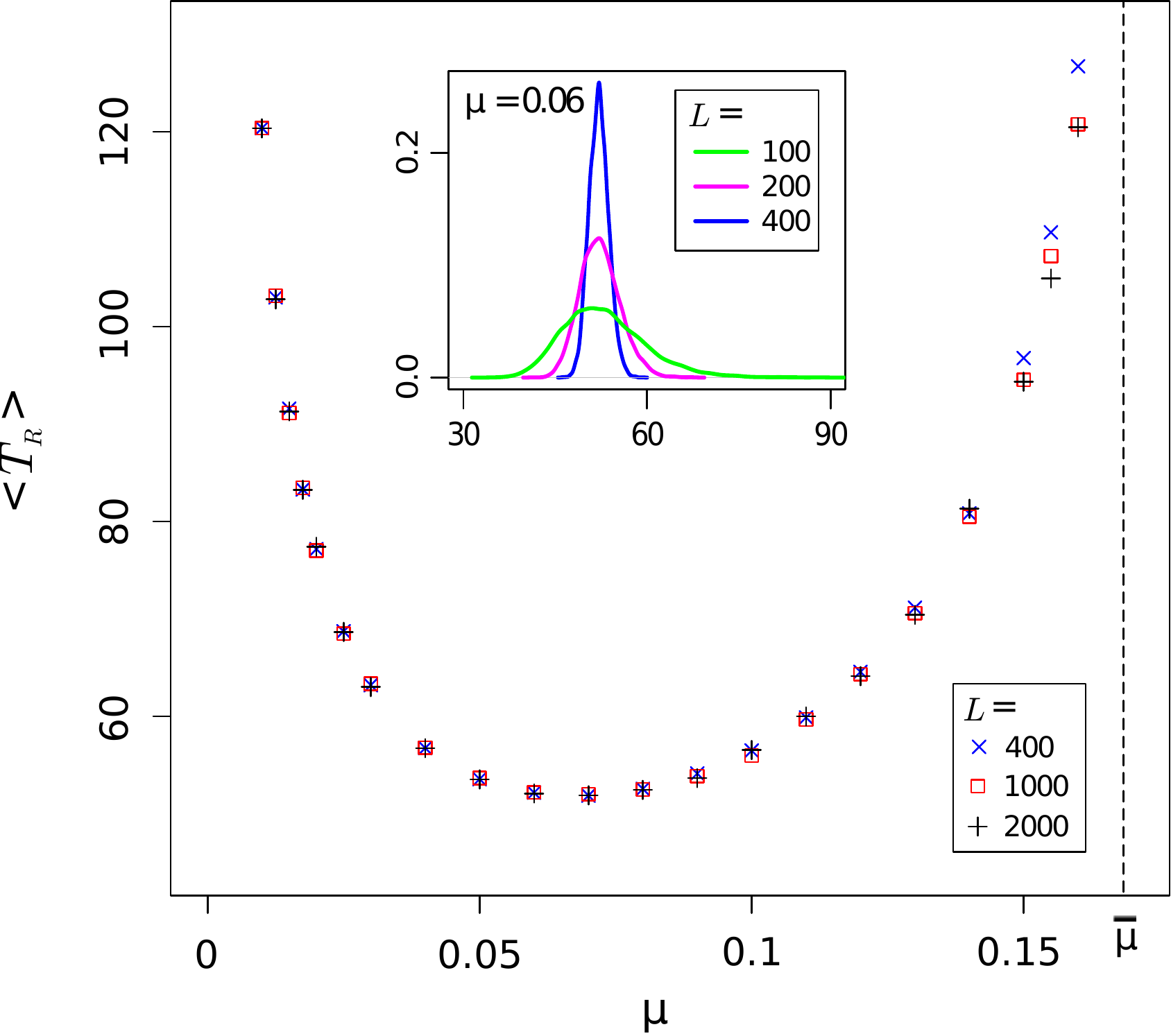}
\caption{(Color online) The mean transition time $\langle T_R\rangle$, necessary for the TM to evolve from complete activation $y(0)=1.0$ to a given intermediate value $y(T_R)=y_*^\infty\approx 0.6752$. The TM is posed on a two-dimensional lattice of size $N=L^2$ with neighborhood $8$ and Gaussian distribution of the thresholds with the mean 0.55 and standard deviation 0.225. The inset illustrates the distribution of $T_R$ at $\mu=0.1$ for $L=100$~(green (light)), $200$~(magenta (medium)) and $400$~(blue (dark))}
\label{f:exit-times}
\end{figure}

For mixing rates $\mu<\bar\mu$, the system will always move towards the lower equilibrium $y_-^\mu$.  We therefore explore the ``transition time'' $T_R$ -- how long it takes to move from the fully activated state $y(0)=1.0$ to some intermediate activation level, chosen to be near, but to the right of, the lower equilibrium.

From simulations, we see that the distribution of transition times becomes narrower for larger $N$, see the inset~\fref{exit-times}. Hence, it is important to look how the mean value $\langle T_R\rangle$ changes for $N\gg1$. It turns out that the dependence $\langle T_R\rangle$ vs. $\mu$ is concave and has a minimum at some intermediate value $\hat\mu$. We see that it becomes arbitrarily large for small mixing rates and exponentially decreases as $\mu$ approaches $\hat\mu$. From the other side the value $\langle T_R\rangle$ increases as $\mu$ becomes closer to $\bar\mu$, see~\fref{exit-times}.

The minimum in the transition-time curve can be explained in terms of two countervailing effects of increased mixing. When the mixing rate is very slow, any changes in behavior take a long time to spread through the lattice. When the mixing rate is high, for our parameters, exchange between neighborhoods tends to preserve the ``upper equilibrium'', leading to an exponentially slow transition on the finite lattice (and no transition at all for the infinite system). Thus, the most rapid transition from the activated state to the lower equilibrium occurs at an intermediate value. 

\section{Conclusions}

Offering an explanation why collective behavior can shift abruptly from avoidance to adoption of an alternative, or \emph{vice versa}, \ct{Granovetter1978} provides an example where a slight change in distribution of individual thresholds leads to completely different outcomes: a system initially at one stable equilibrium switches to another one due to a change of the threshold of one individual. This change can be visualized as a change in the shape of the $F_n$-curve, which we will call the ``activation curve''. We investigate other factors that can change this curve and lead to similar phenomena, including abrupt changes in outcome when an equilibrium disappears.

We model a population on a lattice, with a finite interaction neighborhood, and ``mixing'' -- implemented by exchanging random individuals. To disentangle the effects of neighborhood size and ``locality'', we first considered a lattice with finite neighborhoods in the infinite-mixing limit. We showed that the effect of finite neighborhoods is to ``flatten'' the activation curve, often leading eventually to the elimination of the ``weaker'' equilibrium, as neighborhood size gets smaller.

We then consider the effect of ``locality'', by reducing the mixing rate to be of the same order as the update rate. This system is harder to analyze, but we show that it tends to converge towards a manifold, whose projection can be interpreted in a way similar to the activation curve. This interpretation allows us to define an effective potential in the finite mixing case, which can aid in qualitative analysis. We find that the effect of locality on the projection curve is similar to the effect of finite neighborhood size: it flattens the curve and eventually leads to the disappearance of the weaker equilibrium.

The flattening due to finite neighborhood size \fref{tsp1} can be understood in terms of averaging. If each individual evaluates a random, finite subset of the population when updating, the realized activation curve is a weighted average of the original curve. This averaging tends to flatten out curvature: in the limit of considering a single neighbor, the activation curve becomes a straight line.  Finite mixing has a similar effect~\fref{SM1}.  Individuals' states will be correlated with those of their neighbors, since they are responding to each other.  This increases the variance in neighborhood activation perceived by individuals, for a given value of the mean activation, accentuating the effect of averaging and further smoothing the activation curve.

Here we, as others in the past \ct{Krapivsky2010,Barra2012}, use the mapping between the threshold model and the random field Ising model so that it is possible also to apply tools from statistical physics to the question. Another possibly useful analogy can be made between the TM and a spin gas. We neglect the structure of the network and consider particles stochastically moving in uniform medium, and affected primarily by nearby particles. In this case, the mixing rate can be associated directly with thermodynamic temperature. There is then an analogy between the tendency of all spins to be at the lower equilibrium for small mixing rate in the original system, and low-temperature Bose-Einstein condensation in the spin gas~\ct{Leggett2001}. This mapping may be worth future study.

From sociological point of view, mixing is associated with the rate of information flow in a given society or people mobility. We might imagine ``activists'' who have high mixing rates, and who are eager to change the prevalent behavior. We have seen that large mixing rates can actually prevent the system from switching to a desirable equilibrium, so that an unpopular social norm persists, while low mixing rates facilitate the abandonment of the social norm. However, very low mixing rates make the transition very slow, so that in many cases the transition will happen fastest at intermediate mixing rates.

\section*{Acknowledgments}

Authors were supported by J.S. McDonnell Foundation. This work was made particularly by using the facilities of the Shared Hierarchical Academic Research Computing Network (SHARCNET:www.sharcnet.ca) and Compute/Calcul Canada. Authors are grateful to anonymous reviewers for their remarks, A.R.A. is thankful to Gustavo D\"uring (NYU, USA) and Yevgeni Sh. Mamasakhlisov (Yerevan State University, Armenia) for helpful discussions.

\section*{Appendix A: Ising model framework}

In our model, each node $i \in \{1\ldots N\}$ of the network has a state (or spin) $s_i$ and a (constant) threshold $T_i$. The classical Ising model translates to the a majority rule (MR) model, where all thresholds are exactly $0.5$: a spin tends to flip to the state where it will be aligned with more than one half its neighboring spins, see Ch.~8 \ct{Krapivsky2010}. This system is Hamiltonian with the energy function: $\mathcal H = -\frac12\sum\nolimits_{i;j\in\langle i\rangle} s_is_j$, where $i \in 1\ldots N$, and $\langle i\rangle$ refers to the network neighbors of node $i$.

When the thresholds are randomly distributed with a given probability distribution function (PDF), the model becomes equivalent to the spin system under a magnetic field which describes effects of locality between spins, and such that the strength of nearest interactions depends on the connectivity of the network. In this case, the system also obeys Hamiltonian dynamics and its energy function has the form
\[
\mathcal H = -\sum\limits_{i;j\in\langle i\rangle} \frac{s_is_j}{2n_i} + \sum_i(2T_i-1)s_i\;.
\]
where $n_i$ is the number of connections for a spin $s_i$ and $T_i$ is a given threshold of it. Thus, the induced magnetic field is $h_i = 2T_i-1$.

To simulate the TM dynamics, the following underlying update rule is posed for each update event
\begin{equation}\label{e:UpdateRule}
s_i\mapsto\sign\left(-\frac{\partial\mathcal H}{\partial s_i}\right)=\sign\!\left(\frac1{n_i}\sum\limits_{j\in\langle i\rangle} s_j-h_i\right)\,,
\end{equation}
while the thermodynamic temperature, determining the rate of random flips of spins, is set to zero. Hence, we use only the first part of the Metropolis algorithm~\ct{Glauber1963} that consists only of \eref{UpdateRule} in order to simulate the dynamics of the TM. The second part when the spin might be flipped even if it was not updated due to \eref{UpdateRule} is omitted.

Note that \eref{UpdateRule} indeed allows a sociological interpretation of the TM: if the proportion of plus spins (individuals adopting the social norm) in the neighborhood of a spin $s_i$ is written as $y_i^\circ=\frac1{n_i}\sum\nolimits_{j\in\langle i\rangle}\frac{1+s_j}2$, such that $n_i$ is the connectivity of a spin $s_i$, then \eref{UpdateRule} transforms to the following form: $s_i\mapsto\sign(y_i^\circ-T_i)$. In a particular case of the MR, it translates to the simplest form: $s_i\mapsto\sign\sum\nolimits_{j\in\langle i\rangle} s_j=\sign(y_i^\circ-1/2)$.

\section*{Appendix B: Technical details of simulations}

In our simulations, individuals' initial states and thresholds are independently identically distributed with a given initial activation level $y(0)$ and PDF of thresholds. After that, we initialize the simulation process, using the Metropolis algorithm. At each event, we either update with probability $(1+\mu)^{-1}$ a state of a randomly chosen individual  or exchange with complementary probability the locations of two randomly chosen individuals. We associate the time only with update events, by defining the time quanta $1/N$. We say that the TM is located near the equilibrium point if it fluctuates near it over a sufficiently long period of time, compared with the time of observation.

To construct \fref{tsp4} and \fref{y0_vs_muC}, we considered the TM with $y_-^\infty < y(0) < y_+^\infty$. In this case, we expect the system to move toward either the lower or upper equilibrium.  We run the simulations until they traverse 85\% of the distance from the starting point to one of the mean-field equilibria;  $p_+$ is the probability that it has moved toward the upper equilibrium.


\begin{thebibliography}{99}

\bibitem{Schelling1971} Schelling, T. 1971 Dynamic Models of Segregation. \emph{J. Math. Soc.} \textbf{{1}}, 143--186. (DOI 10.1080/0022250X.1971.9989794)

\bibitem{Granovetter1978} Granovetter, M. S. 1978 Threshold Models of Collective Behavior. \emph{{Amer. J. Soc.}} \textbf{{83}}, 1420--1443 (DOI 10.1086/226707)

\bibitem{Kennedy2009} Kennedy, A. 2009 Mutilation and beautification. \emph{{Austr. Fem. Stud.}} \textbf{{24}}, 211--231 (DOI 10.1080/08164640902852423)
    
\bibitem{WHO2008} World Health Organization 2008 \emph{{Eliminating female genital mutilation. An interagency statement}}.
    
\bibitem{TagEldin2008} Tag-Eldin, M. A., Gadallah, Mo. A., Al-Tayeb, M. N., Abdel-Aty, M., Mansourc, E. \& Sallema, M. 2008 Prevalence of female genital cutting among Egyptian girls. \emph{{Bull. World Health Organ.}} \textbf{{86}}, 269--274 (DOI 10.1590/S0042-96862008000400011)  

\bibitem{UNICEF2010} UNICEF Innocenti Digest 2010 \emph{{The dynamics of social change towards the abandonment of Female Genital Mutilation/Cutting in Five African Countries}}.

\bibitem{Mackie1996} Mackie, G. 1996 Ending footbinding and infibulation: a convention account. \emph{{Amer. Soc. Rev.}} \textbf{{61}}, 999--1017.

\bibitem{Bicchieri2006} Bicchieri, C. 2006 \emph{{The Grammar of Society: The Nature and Dynamics of Social Norms}} Cambridge: Cambridge University Press.
\bibitem{SanMiguel2005} San Miguel, M., Eguiluz, V., Toral, R., \& Klemm, K. 2005 Binary and Multivariate Stochastic models of consensus formation. \emph{{Computing in Science and Engineering}} \textbf{{7}}(6), 67--73 (DOI 10.1109/MCSE.2005.114)

\bibitem{Castellano2009} Castellano, C., Fortunato, S. \& Loreto, V. 2009 Statistical physics of social dynamics. \emph{{Rev. Mod. Phys.}} \textbf{81}, 591--646 (DOI 10.1103/RevModPhys.81.591)

\bibitem{Liggett1999} Liggett, T. M. 1999 \emph{{Stochastic Interacting Systems: Contact, Voter, and Exclusion Processes}}. Springer-Verlag, New York.

\bibitem{Vilone2012} Vilone, D., Ramasco, J. J., Sanchez, A. \& San Miguel, M. 2012 Social and strategic imitation: the way to concensus. \emph{{Scientific Reports}} \textbf{2}, 686 (DOI 10.1038/srep00686)

\bibitem{Krapivsky2003} Krapivsky, P. L. \& Redner, S. 2003 Dynamics of majority rule in two-state interacting spin systems. \emph{{Phys. Rev. Lett.}} \textbf{90}, 238701 (DOI 10.1103/PhysRevLett.90.238701)

\bibitem{Barra2012} Barra, A. \& Agliari, E. 2012 A statistical mechanics approach to Granovetter theory. \emph{Physica A} \textbf{{391}}, 3017--3026. (DOI 10.1016/j.physa.2012.01.007)

\bibitem{Dorogovtsev2008} Dorogovtsev, S. N., Goltsev, A. V. \& Mendes, J. F. F. 2008 Critical phenomena in complex networks. \emph{{Rev. Mod. Phys.}} \textbf{{80}}, 1275--1335. (DOI 10.1103/RevModPhys.80.1275)
    
\bibitem{Sethna1993} Sethna, J. P., Dahmen, K., Kartha, S., Krumhansl, J. A., Roberts, B. W. \& Shore, J. D. 1993 Hysteresis and hierarchies: Dynamics of disorder-driven first-order phase transformations. \emph{{Phys. Rev. Lett.}}, \textbf{70}, 3347--3350 (DOI 10.1103/PhysRevLett.70.3347)
    
\bibitem{Rosinberg2009b} Rosinberg, M. L. \& Munakata, T. 2009 Hysteresis and complexity in the mean-field random-field Ising model: The soft-spin version. \emph{Phys. Rev. B} \textbf{{79}}, 174207 (DOI 10.1103/PhysRevB.79.174207)

\bibitem{Brock2001} Brock, W. A. \& Durlauf, S. N. 2001 Discrete Choice with Social Interactions. \emph{{Rev. Econ. Stud.}} \textbf{68},
235--260. (DOI 10.1111/1467-937X.00168)

\bibitem{Malarz2011} Malarz, K., Korff, R. \& Ku{\l}akowski, K. 2011 Norm breaking in a queue -- athermal phase transition. \emph{{Int. J. Mod. Phys. C}}, \textbf{22}, 719--728 (DOI 10.1142/S0129183111016567)

\bibitem{Levin1974} Levin, S. A. 1974 Dispersion and population interactions. \emph{{Amer. Nat.}} \textbf{{108}}, 207--228.
    
\bibitem{Blasius1999} Blasius, B., Huppert, A. \& Stone, L. 1999 Complex dynamics and phase synchronization in spatially extended ecological systems. \emph{{Nature}} \textbf{{399}}, 354--359. (DOI 10.1038/20676)

\bibitem{Agliari2006} Agliari, E., Burioni, R., Cassi, D. \& Neri F. M. 2006 Efficiency of information spreading in a population of diffusing agents. \emph{{Phys. Rev. E}} \textbf{73}, 046138. (DOI 10.1103/PhysRevE.73.046138)
    
\bibitem{Reichenbach2007} Reichenbach, T., Mobilia, M. \& Frey, E. 2007 Mobility promotes and jeopardizes biodiversity in rock-paper-scissors games. \emph{{Nature}} \textbf{{448}}, 1046--1049 (DOI 10.1038/nature06095)

\bibitem{Watts1998} Watts, D. J. \& Strogatz, S. H. 1998 Collective dynamics of ``small-world'' networks. \emph{{Nature}} \textbf{393}, 440--442 (DOI 10.1038/30918)

\bibitem{Dominicis2006} De Dominicis, C. \& Giardina, I. 2006 \emph{{Random fields and spin glasses: a field theory approach}}. Cambridge: Cambridge University Press.

\bibitem{Krapivsky2010} Krapivsky, P. L., Redner, S. \& Ben-Naim, E. 2010 \emph{{A Kinetic View of Statistical Physics}}. Cambridge: Cambridge University Press.

\bibitem{Bolker1999} Bolker, B.M. and Pacala, S.W. 1999 Using moment equations to understand stochastically driven spatial pattern formation in ecological systems. \emph{{Th. Pop. Bio.}} \textbf{{52}}, 179--197. (DOI 10.1006/tpbi.1997.1331)
    
\bibitem{Murrell2004} Murrell, D. J., Dieckmann, U. \& Law, R. 2004 On moment closures for population dynamics in continuous space. \emph{{J. Theor. Biol.}} \textbf{{229}}, 421--432 (DOI 10.1016/j.jtbi.2004.04.013)

\bibitem{Murrell2009} Murrell, D. J. 2009 On the emergent spatial structure of size-structured populations: when does self-thinning lead to a reduction in clustering? \emph{{J. Ecol.}}, \textbf{{97}}, 256--266. (DOI 10.1111/j.1365-2745.2008.01475.x)
    
\bibitem{Toral2007} Toral, R. \& Tessone, C. J. 2007 Finite size effects in the dynamics of opinion formation. \emph{{Comm. Comp. Phys.}} \textbf{{2}}, 177--195.

\bibitem{Gardiner2004} Gardiner, C. W. 2004 \emph{{Handbook of Stochastic Methods}}, 3rd edn. Springer.
    
\bibitem{Leggett2001} Leggett, A. J. 2001 Bose-Einstein condensation in the alkali gases: Some fundamental concepts. \emph{{Rev. Mod. Phys.}} \textbf{73}, 307--356. (DOI 10.1103/RevModPhys.73.307)
    
\bibitem{Glauber1963} Glauber, R. J. 1963 Time-Dependent Statistics of the Ising Model. \emph{{J. Math. Phys.}} \textbf{{4}}, 294--308 (DOI 10.1063/1.1703954)

\end{thebibliography}
\end{document}